\documentclass[floatfix,twocolumn,showpacs,preprintnumbers,amsmath,amssymb,pra,superscriptaddress,longbibliography]{revtex4-1}
\usepackage{color}
\usepackage[usenames,dvipsnames,svgnames,table]{xcolor}
\usepackage[colorlinks=true,linkcolor=blue,urlcolor=blue,citecolor=blue]{hyperref}

\usepackage{mathtools}
\usepackage{graphicx}
\usepackage{dcolumn}
\usepackage{array}
\usepackage{lipsum}
\usepackage{bm}
\usepackage{subfigure}
\usepackage{amssymb}
\usepackage{multirow}
\usepackage{tabularx}
\usepackage{amsmath}
\usepackage{braket}

\usepackage{enumitem}
\renewcommand{\vec}[1]{\mathbf{#1}}

\begin{document}

%\title{Local Field Correction of Correlated Electrons From KS-DFT}

\title{ \emph{Ab initio} Static Exchange--Correlation Kernel across Jacob’s Ladder without functional derivatives}

% \title{ \emph{Ab initio} Static Exchange--Correlation Kernel of Real Materials across Jacob’s Ladder without functional derivatives}

% \title{\emph{Ab initio} Exchange--Correlation Kernel from Density Functional Theory Simulations: From Ambient Conditions to Warm Dense Matter}

\author{Zhandos A. Moldabekov}
\email{z.moldabekov@hzdr.de}
\affiliation{Center for Advanced Systems Understanding (CASUS), D-02826 G\"orlitz, Germany}
\affiliation{Helmholtz-Zentrum Dresden-Rossendorf (HZDR), D-01328 Dresden, Germany}

\author{Maximilian B\"ohme}
\affiliation{Center for Advanced Systems Understanding (CASUS), D-02826 G\"orlitz, Germany}
\affiliation{Helmholtz-Zentrum Dresden-Rossendorf (HZDR), D-01328 Dresden, Germany}

\affiliation{Technische  Universit\"at  Dresden,  D-01062  Dresden,  Germany}

\author{Jan Vorberger}
\affiliation{Helmholtz-Zentrum Dresden-Rossendorf (HZDR), D-01328 Dresden, Germany}

\author{David Blaschke}
\affiliation{Institute of Theoretical Physics, University of Wroclaw, 50-204 Wroclaw, Poland}

\author{Tobias Dornheim}
\email{t.dornheim@hzdr.de}
\affiliation{Center for Advanced Systems Understanding (CASUS), D-02826 G\"orlitz, Germany}
\affiliation{Helmholtz-Zentrum Dresden-Rossendorf (HZDR), D-01328 Dresden, Germany}

%%%%%%%%%%%%%%%%%%%%%%%%%%%%%%%%%%%%%%%%%%%%%%%%%%%%%%%%%%%%%%%%%%%%%%%%%%%%%%%%
% Abstract
%%%%%%%%%%%%%%%%%%%%%%%%%%%%%%%%%%%%%%%%%%%%%%%%%%%%%%%%%%%%%%%%%%%%%%%%%%%%%%%%
\begin{abstract}
The electronic exchange—correlation (XC) kernel constitutes a fundamental input for the estimation of a gamut of material properties such as the dielectric characteristics, the thermal and electrical conductivity, or the response to an external perturbation. In practice, no reliable method has been known that allows to compute the kernel of real materials with arbitrary XC functionals. In this work, we overcome this long-standing limitation by introducing a new, formally exact methodology for the computation of the material specific static XC kernel exclusively within the framework of density functional theory (DFT)  and without employing functional derivatives---no external input apart from the usual XC-functional is required. We compare our new results with exact quantum Monte Carlo (QMC) data for the archetypical uniform electron gas model at both ambient and warm dense matter conditions. This gives us unprecedented insights into the performance of different XC-functionals, and has important implications for the development of new functionals that are designed for the application at extreme temperatures. In addition, we obtain new DFT results for the XC kernel of warm dense hydrogen as it occurs in fusion applications and astrophysical objects. The observed excellent agreement to the QMC reference data demonstrates that our framework is capable to capture nontrivial effects such as XC-induced isotropy breaking in the density response of hydrogen at large wave numbers.

\end{abstract}

\maketitle

%%%%%%%%%%%%%%%%%%%%%%%%%%%%%%%%%%%%%%%%%%%%%%%%%%%%%%%%%%%%%%%%%%%%%%%%%%%%%%%%
% Motivation and Introduction
%%%%%%%%%%%%%%%%%%%%%%%%%%%%%%%%%%%%%%%%%%%%%%%%%%%%%%%%%%%%%%%%%%%%%%%%%%%%%%%%
% \section{Introduction}

The density functional theory (DFT) approach~\cite{Jones_RMP_1989,Jones_RevModPhys_2015} is arguably the most successful simulation tool in many-body physics, quantum chemistry,  and related disciplines. Its main advantage is the evened out balance between reasonable accuracy and manageable computation cost, which allows for the \emph{ab initio} description of real materials. While formally exact~\cite{Hohenberg_Kohn_1964}, DFT requires as external input the a-priori unknown exchange--correlation (XC) functional, which, in practice, has to be approximated. At ambient conditions, where the electrons are in the ground state, Jacob's ladder of functionals~\cite{doi:10.1063/1.1390175,PhysRevLett.91.146401} serves as a useful categorization of different approximations, and 
% Furthermore, there exists an extensive body of literature on the empirical benchmark of a gamut of functionals for various applications~\cite{Clay_PRB_2016, doi:10.1063/1.5017198, Borlido2020, doi:10.1021/acs.jctc.9b00322, Moldabekov_PRB_2022, Moldabekov_JCP_2021}.
%%%Indeed, 
the number of publications that utilize the DFT approach has been exponentially increasing over the last years~\cite{doi:10.1146/annurev-physchem-040214-121420}. %%% Particularly,  the DFT method become a standard tool 
%%%for the investigation of the warm dense matter (WDM) which is routinely generated by laser-induced shock compression \cite{BH16}, exists
%%%in planetary interiors \cite{Benuzzi_Mounaix_2014, Kraus2017, Lazicki2021}, and occurs in Inertial Confinement Fusion Implosions \cite{PhysRevLett.104.235003}. 

The drastic reduction of the computation cost that renders DFT simulations feasible is achieved by a formally exact mapping %%%the original many-electron problem  
onto an effective single-electron problem~\cite{Hohenberg_Kohn_1964}. %%% While, in theory, being exact~\cite{Hohenberg_Kohn_1964}, 
Unfortunately, the bulk of information about electron--electron correlations is lost in the process, and DFT gives straightforward access only to the single-electron density $n_e(\mathbf{r})$ and different contributions to the energy in practice; electron--electron correlations cannot be readily estimated. %%% This is a highly unsatisfactory situation as many potentially profound insights into the system of interest are lost. 
%%%In practice,
Therefore, advanced DFT applications such as linear-response time-dependent DFT (LR-TDDFT)~\cite{PhysRevLett.76.1212} %%%, being formally exact framework for calculating electronic excitations \cite{PhysRevLett.76.1212}, 
require as an additional input the material specific XC-kernel~\cite{Goerling_PRB_2019,Goerling_PRA_1998} $K_\textnormal{xc}(\mathbf{q},\omega)$. Yet, very little is known about the actual XC-kernel of real materials~\cite{Boehme_PRL_2022,Boehme_Folgepaper}, and hitherto no feasible and universal way to compute it has been known. %%%Notably, previous XC kernels had not been consistent to the XC \textit{potential} of Kohn–Sham DFT calculations for extended systems~\cite{Byun_2020}. %%%The situation  is even worse at finite (extreme) temperatures, where little is known about thermal XC kernel beyond UEG .
In particular, there had been no possibility to compute the XC kernel for the existing great variety of XC functionals (more than 400) beyond the adiabatic LDA (ALDA) and GGA (AGGA) for extended systems~\cite{Byun_2020}.%%%  with sampling points in the first Brillouin zone (k point sampling).

% The LR-TDDFT   the KS (non-interacting) response function and XC kernel. The KS response function is constructed from KS wave-functions of the equilibrium (ground) state and XC kernel is computed separately in a certain approximation.  

In this Letter, we overcome these fundamental limitations
by introducing a new, formally exact methodology for
the \emph{ab initio} calculation of the non-local static XC-kernel within the framework of DFT; it is fully compatible with the XC potential of selfconsistent Kohn–Sham (KS) equilibrium calculations for any XC functional. 
The presented approach completely circumvents the problem of computing functional derivatives, which had been the key obstacle that prevented going beyond AGGA for extended systems. 

Specifically, we propose to use standard KS-DFT to compute the single-particle density $n_e(\mathbf{r})$ for a given electronic Hamiltonian $\hat{H}_e$. As a second step, we repeat the calculation for a modified Hamiltonian $\hat{H}_{\mathbf{q},A}=\hat{H}_e + \hat{V}_\textnormal{ext}(\mathbf{q},A)$ that is subject to a monochromatic external perturbation $\hat{V}_\textnormal{ext}(\mathbf{q},A)=2 A \sum_{j=1}^N \textnormal{cos}\left( \mathbf{q}\cdot\mathbf{r}_j \right)$ of wave vector $\mathbf{q}$ and perturbation amplitude $A$~\cite{moroni,moroni2,Dornheim_PRL_2020}; this gives us the perturbed single-particle density $n_e(\mathbf{r})_{\mathbf{q},A}$. We can thus compute the induced density modulation as $\Delta n_e(\mathbf{r})_{\mathbf{q},A}=n_e(\mathbf{r})_{\mathbf{q},A} - n_e(\mathbf{r})$. %%% due to the external perturbation. Third, i
In the limit of small $A$, the latter directly gives us the static linear density response function $\chi(\mathbf{q})=\Delta n_e(\mathbf{r})_{\mathbf{q},A}/2A\textnormal{cos}\left( \mathbf{q}\cdot\mathbf{r} \right)$, which---in the general dynamic (i.e., $\omega\neq0$) case---can be expressed for a homogeneous system (see the Supplemental Material~\cite{supplement} for more details) as~\cite{kugler1,quantum_theory}
\begin{eqnarray}\label{eq:kernel}
 \chi(\mathbf{q},\omega) = \frac{\chi_0(\mathbf{q},\omega)}{1 - \left[v(q)
 +K_\textnormal{xc}(\mathbf{q},\omega)\right]\chi_0(\mathbf{q},\omega)}\ .
\end{eqnarray}
Here $v(q)=4\pi/q^2$ is the Coulomb interaction, and $\chi_0(\mathbf{q},\omega)$ denotes a known reference function, such as the Lindhard function in the case of a uniform electron gas (UEG)~\cite{quantum_theory}. 
In this case, the XC-kernel,  too, has a well-defined physical meaning and contains the full wave-vector- and frequency-resolved information about electronic XC-effects in the system. 
Moreover, it is then directly related to the \emph{local field correction} $G(\mathbf{q},\omega)= - K_\textnormal{xc}(\mathbf{q},\omega)/v(q)$ that is the central property within dielectric theories~\cite{stls_original,stls,vs_original,dynamic_ii,Tolias_JCP_2021,tanaka_hnc}.
Hence, setting $K_\textnormal{xc}\equiv0$  leads to the mean-field level description, which is commonly known as the \emph{random phase approximation} (RPA)~\cite{quantum_theory}, $\chi_\textnormal{RPA}(\mathbf{q},\omega)$.
 In the case of an inhomogeneous electron gas, for example in the potential of a fixed ion configuration, it is common practice to use the KS response function $\chi_\textnormal{KS}(\mathbf{q}, \omega)$ as $\chi_0(\mathbf{q}, \omega)$ ~\cite{ullrich2012time,marques2012fundamentals}.
As the final step, we invert Eq.~(\ref{eq:kernel}) to compute the static XC-kernel $K_\textnormal{xc}(\mathbf{q})$ for any given system.

%, or the KS response function~\cite{ullrich2012time,marques2012fundamentals} $\chi_\textnormal{KS}(\mathbf{q},\omega)$ in the case of an inhomogeneous system. Having both $\chi(\mathbf{q},0)$ (from our new framework) and $\chi_0(\mathbf{q},0)$, it is straightforward to invert Eq.~(\ref{eq:kernel}) for the consistent, material-specific static XC-kernel $K_\textnormal{xc}(\mathbf{q},0)$.

%%%\textcolor{red}{mark}
%%%The latter in combination with the KS response function $\chi_0=\chi_\textnormal{KS}(\mathbf{q}, 0)$ defines the static XC-kernel $K_\textnormal{xc}(\mathbf{q})=K_\textnormal{xc}(\mathbf{q},0)$ via inversion of Eq. (\ref{eq:kernel}),
% \begin{eqnarray}\label{eq:invert}
%  K_\textnormal{xc}(\mathbf{q}) &=& -\left\{
%  v(q) + \left( \frac{1}{\chi(\mathbf{q})} - \frac{1}{\chi_0(\mathbf{q})} \right)
%  \right\}\ ,
% %  , \\\nonumber
% %  &=& \frac{1}{\chi_\textnormal{RPA}(\mathbf{q})} - \frac{1}{\chi(\mathbf{q})}\ .
% \end{eqnarray}
%%%where $\chi_\textnormal{KS}$ is computed using the KS-orbitals and corresponding energy eigenvalues of the unperturbed equilibrium system ~\cite{ullrich2012time,marques2012fundamentals}.

To rigorously demonstrate the correctness and utility of our new approach, we consider two representative systems. The UEG~\cite{loos,quantum_theory,dornheim_physrep18_0} constitutes the archetypical electronic system and is the basis for many applications such as the BCS theory of superconductivity~\cite{Bardeen_PhysRev_1957} and Fermi liquid theory~\cite{quantum_theory}. % In the context of the present work, the UEG has the considerable advantage that reliable benchmark data for a number of properties are available based on highly accurate quantum Monte Carlo (QMC) calculations~\cite{Foulkes_RevModPhys_2001,dornheim_POP}. 
As a second, even more challenging example, we consider hydrogen, the most abundant element in our universe, which is the subject of active investigation~\cite{RevModPhys.84.1607,Pierleoni_PNAS_2016,Dias_Silvera_Science_2017,doi:10.1126/science.aat0970,Boehme_PRL_2022}. 
% Indeed, many fundamental questions about hydrogen such as the precise nature and location of the insulator-to-metal phase transition~\cite{Pierleoni_PNAS_2016} remain unanswered. 
%Here, we use our new methodology to obtain the static XC-kernel of hydrogen and find very good agreement to the recent exact QMC results by B\"ohme \emph{et al.}~\cite{Boehme_PRL_2022}.
We find excellent agreement with existing highly accurate quantum Monte Carlo (QMC) results~\cite{moroni2,Boehme_PRL_2022,dornheim_ML} in both cases.

In addition, we analyze both the electronic ground state and so-called \emph{warm dense matter} (WDM) at the electronic Fermi temperature, $\theta=k_\textnormal{B}T/E_\textnormal{F}=1$ (with $E_\textnormal{F}$ being the usual Fermi energy~\cite{quantum_theory}). In fact, WDM is ubiquitous in nature~\cite{fortov_review}, and occurs in astrophysical objects such as giant planet interiors~\cite{Benuzzi_Mounaix_2014} and brown dwarfs~\cite{becker}. Moreover, WDM is highly relevant for technological applications such as inertial confinement fusion~\cite{hu_ICF} and the discovery of novel materials~\cite{Lazicki2021,Kraus2017,Kraus2016}. It is well-known that the theoretical description of WDM is notoriously difficult; from the perspective of DFT, one requires a density functional of the XC-free energy $F_\textnormal{xc}$ that explicitly depends on the electronic temperature $T$~\cite{mermin_65,karasiev_importance,kushal}. While first developments~\cite{ksdt,groth_prl17,Karasiev_PRL_2018,Karasiev_PRB_2020} have recently become available, the field of finite-$T$ XC-functionals still remains in its infancy, and the performance of various approximations~\cite{Moldabekov_JCP_2021,Moldabekov_PRB_2022} is substantially less understood compared to $T=0$. Indeed, most DFT calculations for WDM are carried out on the basis of the \emph{zero-temperature approximation} where the XC-free energy is approximated by the XC-energy in the ground state.

The main present limitation of our new approach is given by its restriction to compute the XC-kernel in the limit of $\omega=0$. 
Still, it is possible
to compute the dynamic density response function within the \emph{static approximation}~\cite{dornheim_dynamic}, i.e., by setting $K_\textnormal{xc}(\mathbf{q},\omega)\equiv K_\textnormal{xc}(\mathbf{q})$ in Eq.~(\ref{eq:kernel}).  %$\chi_\textnormal{stat}(\mathbf{q},\omega) = \chi_0(\mathbf{q},\omega)/\left(1 - \left[v(q) + K_\textnormal{xc}(\mathbf{q})\right]\chi_0(\mathbf{q},\omega)\right)$,
%where the dynamic XC-kernel $K_\textnormal{xc}(\mathbf{q},\omega)$ is approximated by its exact static limit. 
In this way, one combines a dynamic description on the level of the RPA with exact static correlations. This approximation has been shown to be highly accurate in the case of the UEG for weak to moderate coupling strengths, including the important regime of metallic densities $r_s\lesssim5$ (with $r_s$ being the Wigner-Seitz radius in Hartree atomic units~\cite{Ott2018}).

\begin{figure}
    \centering
    \includegraphics[width=7 cm]{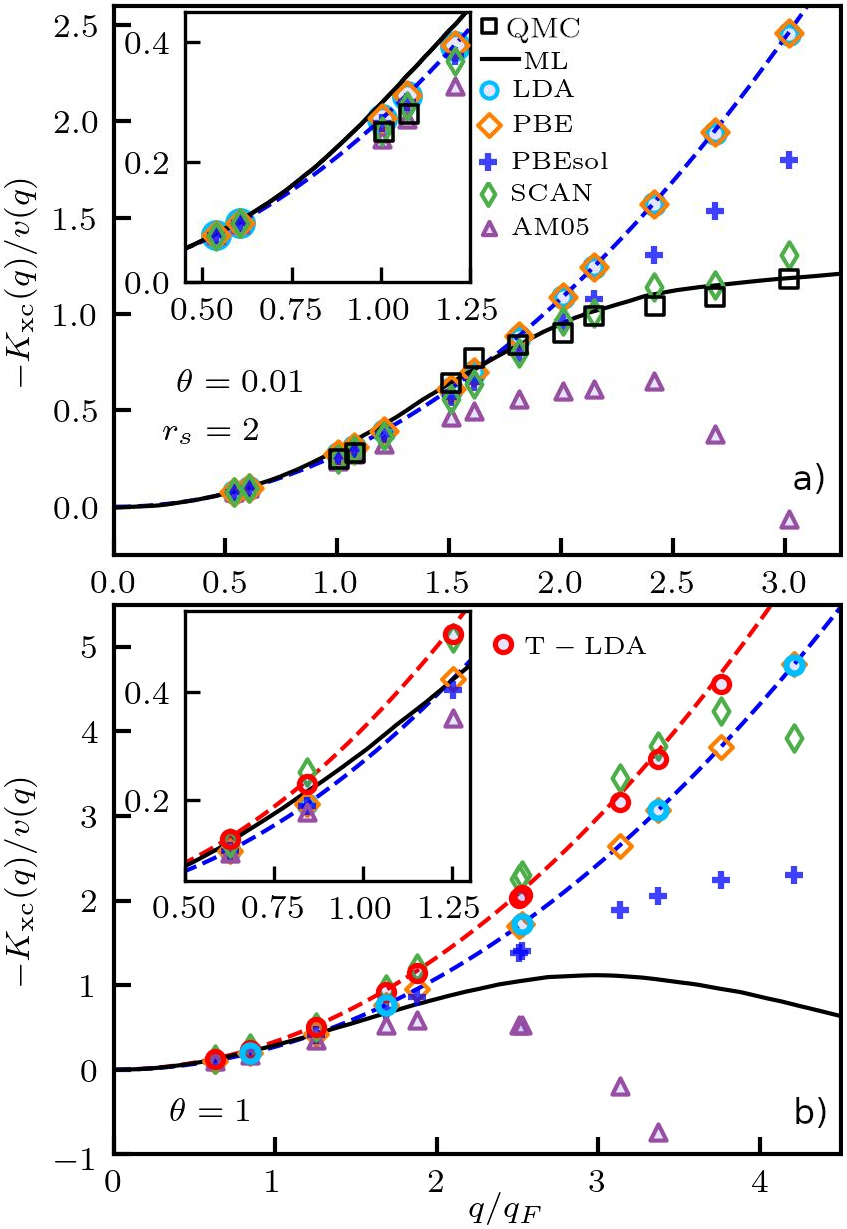}
    \caption{XC-kernel $K_\textnormal{xc}(\mathbf{q})$ of the UEG at a) ambient conditions with  $\theta=0.01$  and  at b) WDM~\cite{new_POP,wdm_book} conditions with  $\theta=1$  for $r_s=2$. Solid black line: exact UEG results based on the neural-net representation of Ref.~\cite{dornheim_ML}. Black squares: exact QMC results~\cite{moroni2}.
    The other symbols distinguish DFT calculations for the density modulation using different XC-functionals.}
    \label{fig:chi_T0}
\end{figure}

\textbf{Results.}
%We begin with an analysis of the XC kernel of the UEG at ambient conditions (i.e., at $T=0$) shown in the top panel of Fig.~\ref{fig:chi_T0}. 
In Fig.~\ref{fig:chi_T0}a), we show the XC kernel of the UEG at $T=0$ for the metallic density of $r_s=2$.
Specifically, we have carried out DFT calculations governed by the perturbed Hamiltonian for multiple wave vectors $\mathbf{q}$ and a sufficiently small perturbation amplitude $A$; the different symbols show results for a selection of popular XC-functionals. In addition, the solid black line corresponds to a parametrization~\cite{dornheim_ML} of the highly accurate QMC results by Moroni \emph{et al.}~\cite{moroni2} (black squares). % and taken from the neural-net representation from Ref.~\cite{dornheim_ML}.
%Fig.~\ref{fig:chi_T0} has been obtained for $r_s=2$, which is a metallic density that can be probed in experiments for example with aluminum~\cite{PhysRevB.40.10181}. 
%To get a more rigorous insight into the performance of the different functionals, we show the corresponding XC-kernel $K_\textnormal{xc}(\mathbf{q})$ that we have obtained by inversion of Eq.~(\ref{eq:kernel}) in the top panel of  Fig.~\ref{fig:chi_T0}.
Throughout this work, we follow the usual convention~\cite{IIT_1987,cdop,dornheim_ML} and divide $K_\textnormal{xc}$ by the Coulomb interaction $v(q)$, resulting in the commonly analyzed local field correction.
In the limit of small $q$, the LFC is know to satisfy the exact compressibility sum-rule~\cite{IIT_1987}, $\lim_{q\to0}G(q) = - \frac{q^2}{4\pi} {\partial^2} \left( n F_\textnormal{xc} \right)/{\partial n^2}$,
with $n=N/V$ being the average number density.
It is depicted as the dashed blue parabola in Fig.~\ref{fig:chi_T0}a); we note that it holds $\lim_{T\to0}F_\textnormal{xc} = E_\textnormal{xc}$.
Evidently, $\lim_{q\to0}G(q)$ is accurately reproduced both by the neural-net representation~\cite{dornheim_ML} and by all depicted XC-functionals for small $q$.
Moreover, both the LDA functional by Perdew and Wang~\cite{LDA_PW} (light blue circles) and the generalized gradient approximation (GGA) by Perdew, Burke and Ernzerhof (PBE~\cite{PBE}, orange diamonds)  have been constructed to reproduce $\lim_{q\to0}G(q)$ for all $q$~\cite{Perdew_Langreth_PRL_1977} in the case of the UEG. This is substantiated by our empirical results and the LDA and PBE give indistinguishable results. % This is expected as all gradient terms vanish in the case of a UEG in PBE by design. 
Remarkably, the parabolic small-$q$ expansion accurately reproduces the QMC data for $q\lesssim2q_\textnormal{F}$; this is a nontrivial observation which explains the success of both the simple LDA and the more sophisticated PBE in the description of bulk materials~\cite{moroni2}. 
In contrast, the AM05 functional by Armiento and Mattson~\cite{PhysRevB.72.085108} (purple up-triangles), which is a semi-local GGA and has been shown to give comparable quality to hybrid functionals in the description of solids \cite{Mattsson}, only reproduces the QMC data for $q\lesssim1.2q_\textnormal{F}$. For large $q$, it has been designed to reproduce the Airy gas model~\cite{Mattsson}, resulting in a substantial drop towards negative values.
The semi-empirical PBEsol~\cite{PBEsol} (blue plusses), on the other hand, is virtually indistinguishable from PBE for $q\lesssim2q_\textnormal{F}$, and exhibits a somewhat higher accuracy at large wavenumbers. Finally, the meta-GGA SCAN~\cite{SCAN} (green diamonds) constitutes by far the most accurate functional for the ground state and gives basically exact results over the entire depicted $q$-range, as it was designed to reproduce the QMC results~\cite{SCAN}.

\begin{figure}
\center
\includegraphics[width=7.5cm]{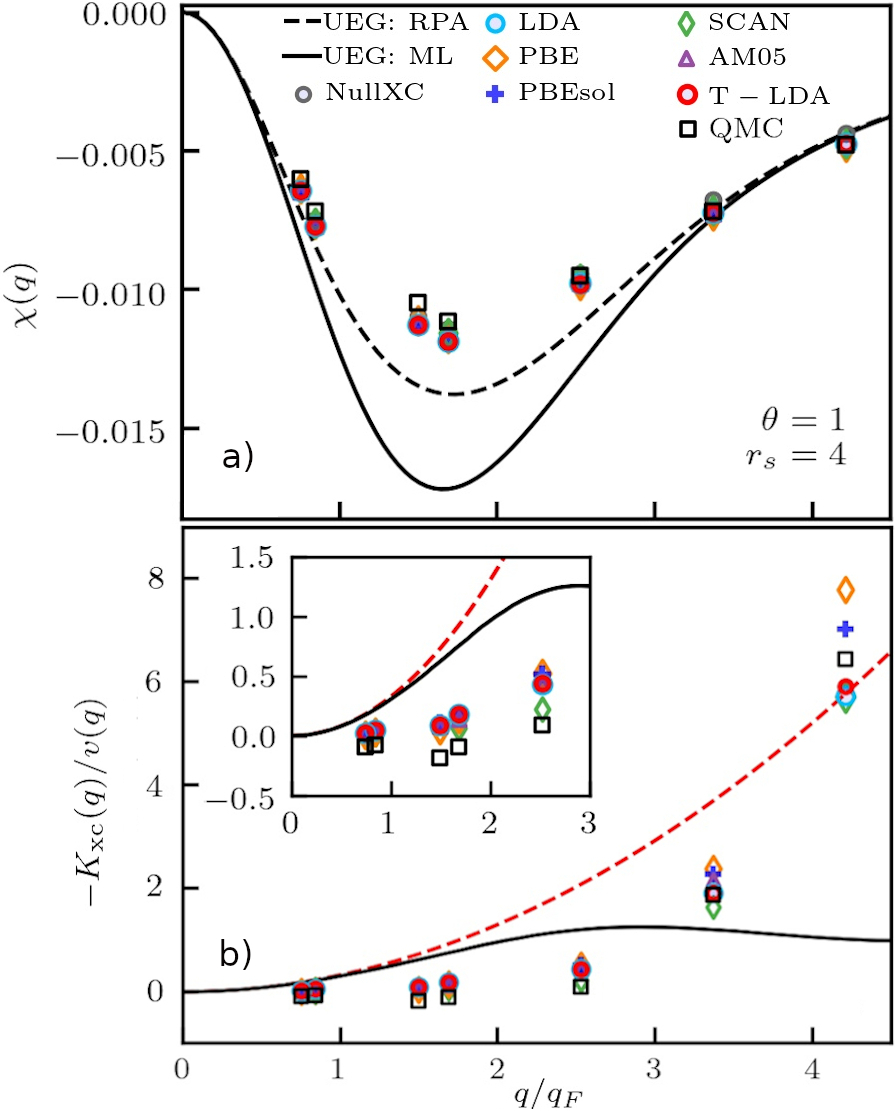}
\caption{ \label{fig:H_rs2_4} 
a) Electronic static density response function $\chi(\mathbf{q})$  and b) XC-kernel $K_\textnormal{xc}(\mathbf{q})$ (bottom panel) 
% computed from the same mean-field reference function (see the main text) $\chi_0(\mathbf{q})$  
of hydrogen at WDM~\cite{new_POP,wdm_book} conditions ($\theta=1$) for  $r_s=4$. Solid (dashed) black line: exact results for the UEG model at the same conditions based on the neural-net representation of Ref.~\cite{dornheim_ML} (analytical RPA). Black squares: exact QMC results for hydrogen~\cite{Boehme_PRL_2022}.
    The other symbols distinguish DFT calculations for the density modulation using different XC-functionals.
}
\end{figure}

In Fig.~\ref{fig:chi_T0}b), we present results for the UEG in the WDM regime, i.e., at the electronic Fermi temperature $\theta=1$.
In this case, we additionally consider the finite-$T$ LDA functional by Groth \emph{et al.}~\cite{groth_prl17} (red circles).
The dashed red and blue lines show the exact small-$q$ expansion $\lim_{q\to0}G(q)$ evaluated at $\theta=1$ and $\theta=0$, respectively. Evidently, the ground-state LDA (and PBE) follows the latter curve, as it is expected. Similarly, the finite-$T$ LDA follows the red curve and, therefore, reproduces the correct impact of the temperature on the small-$q$ limit. This can be seen particularly well in the inset where we show a magnified segment. It can be expected that the recent finite-$T$ GGA functional~\cite{Karasiev_PRL_2018} exhibits the same behavior as it has been constructed to reproduce the finite-$T$ LDA for the UEG.
In practice, however, the impact of $K_\textnormal{xc}(\mathbf{q})$ on the density response function vanishes for $q\to 0$, and even the RPA becomes exact. For $q\gtrsim q_\textnormal{F}$, where the impact of the XC-kernel is most pronounced, the ground-state evaluation of $\lim_{q\to0}G(q)$ coincidentally constitutes a superior approximation to the true curve (solid black). Therefore, the ground-state LDA exhibits a better accuracy than the consistently temperature-dependent functional.
This is a very important point for which a more detailed analysis for $1\leq r_s \leq 6$ and $0\leq \theta \leq 4$ is presented in the Supplementary Material~\cite{supplement}.

We stress that these findings have profound consequences for the construction of the next generation of XC-functionals that are specifically designed for the application at WDM conditions. Evidently, translating Jacob's ladder of functional approximations~\cite{doi:10.1063/1.1390175} from the ground-state to finite temperatures does not necessarily improve the quality of DFT simulations in the WDM regime. Making the lowest rung---i.e., the LDA---explicitly $T$-dependent might actually lead to a deterioration of the attained accuracy. Moreover, this deficiency is, by design, not removed on the GGA-level, which is based on the same $q\to0$ expansion. 

Returning to Fig.~\ref{fig:chi_T0}b), we  find that the ground-state SCAN functional performs similarly poorly as AM05, which is in stark contrast to its impressive accuracy at $T=0$. We thus conclude that the meta-GGA corrections on which SCAN is based strongly depend on the electronic temperature. 
In contrast, the semi-empirical PBEsol provides a more accurate description of the static XC kernel at $q\lesssim2.5q_\textnormal{F}$ than LDA, T-LDA, AM05, and SCAN.

%To demonstrate the broad utility of our new approach, we
Let us next consider hydrogen at extreme conditions---a  state of matter that plays a central role in the description of the implosion path of a fuel capsule towards nuclear fusion~\cite{hu_ICF} and naturally occurs within astrophysical objects such as giant planet interiors~\cite{Benuzzi_Mounaix_2014}.
In Fig.~\ref{fig:H_rs2_4}, we show our new DFT results for the static density response of hydrogen that has been computed for a single fixed ion snapshot from a corresponding DFT-MD simulation. We note that, while the averaging over many snapshots is straightforward, benchmarking DFT for a single proton configuration constitutes an even more rigorous test of our methodology as, in this way, error cancellation between different snapshots is ruled out. 

At $r_s=2$, where hydrogen is known to be mostly ionized, the bulk of the electrons can be categorized as \emph{unbound}, meaning that they are not primarily localized around the protons. Therefore, the density response closely resembles the UEG model and we find the same conclusions as in Fig.~\ref{fig:chi_T0}a), see the Supplemental Material~\cite{supplement}. 

From a physical perspective, the case of $r_s=4$ shown in Fig.~\ref{fig:H_rs2_4} is more interesting.  In addition to the more pronounced impact of Coulomb correlations, hydrogen is partially ionized at these conditions, with an approximate fraction of \emph{free electrons} of $\alpha=0.54-0.6$~\cite{Militzer_Ceperley_PRE_2001,Boehme_PRL_2022}. Consequently, the numerical results for $\chi(\mathbf{q})$, Fig.~\ref{fig:H_rs2_4}a), exhibit a substantially reduced density response compared to the UEG, as the \emph{bound} electrons cannot react as much as the free electrons to the external potential. 
Overall, we find good qualitative agreement between DFT and the QMC data over the entire depicted $q$-range, even though the true reduction of the density response due to the localization around the protons is somewhat underestimated for $q_\textnormal{F}\lesssim q \lesssim3q_\textnormal{F}$. Remarkably, we find that all XC-functionals reproduce the nontrivial increase in the magnitude of $\chi(\mathbf{q})$ compared to the UEG model around $q\sim4q_\textnormal{F}$, which has been explained as a consequence of isotropy breaking in the presence of the proton configuration in Ref.~\cite{Boehme_PRL_2022}.

In Fig.~\ref{fig:H_rs2_4}b), we show the corresponding XC-kernels, that we have extracted from the different $\chi(\mathbf{q})$ data sets via Eq.~(\ref{eq:kernel}). As the reference function, we use a mean-field response function $\chi_0(\mathbf{q})$ that we have obtained from a separate DFT simulation with the XC-functional being set to zero. This has the advantage that XC-kernels from different theories are directly comparable to each other. For completeness, we note that extracting the actual XC-functional dependent kernel by inserting the respective $\chi_\textnormal{KS}(\mathbf{q},0)$ into Eq.~(\ref{eq:kernel}) is straightforward, but would make the direct comparison less meaningful.
The resulting data for $K_\textnormal{xc}(\mathbf{q})$ of hydrogen at $r_s=4$ and $\theta=1$ qualitatively agree with each other, but starkly disagree from the UEG model at these conditions. In particular, the kernel attains remarkably small values for $q\lesssim2.5q_\textnormal{F}$, followed by a pronounced increase for $q\gtrsim3q_\textnormal{F}$. Clearly, our new methodology is capable to accurately capture the complex interplay of the ion structure with electronic XC-effects as they manifest in $K_\textnormal{xc}(\mathbf{q})$.
In addition, we do not find the simple parabolic behaviour for $\lim_{q\to0}G(q)$ from LDA.
% Therefore, the commonly used ALDA is an approximation to the true material specific LDA XC kernel. 
% To understand this, we recall that both ALDA and AGGA are defined and computed locally utilizing $\delta(\vec r -\vec r^{\prime})$ \cite{PhysRevB.86.081103, PhysRevB.99.035151}.
In fact, $K_{\rm xc}(\vec q)$ from the ALDA  always reduces to a constant, leading to $G(q)\sim q^{2}$ \cite{Byun_2020, PhysRevB.35.5585}. In our method, the total response of a system to a harmonic perturbation with a given wave vector is selfconsistently determined by the electron density everywhere in the simulation cell.

\begin{figure}[t!]
\center
\includegraphics[width=7cm]{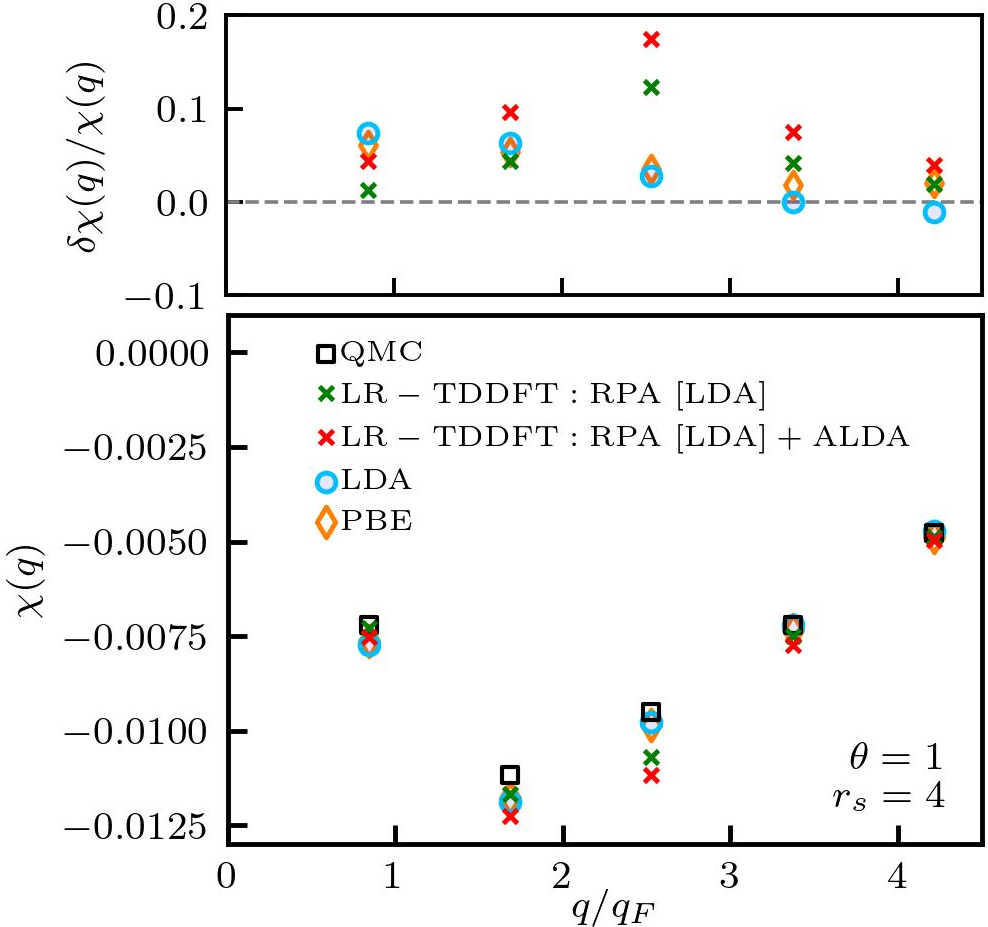}
\caption{ \label{fig:H_LDA_rs4} 
Illustration of the inconsistent combination of $\chi_\textnormal{KS}(\mathbf{q},\omega)$ with the ALDA kernel. Bottom:
Static density  response of warm dense hydrogen with $r_s=4$ and $\theta=1$; the light blue circles, orange diamonds, and black squares are taken from Fig.~\ref{fig:H_rs2_4}. Green crosses: Static ($\omega\to0$) limit of LR-TDDFT results within RPA  based on $\chi_\textnormal{KS}(\mathbf{q},\omega)$ using KS-orbitals from a ground-state LDA calculation. Red crosses: corresponding ALDA results. Top panel: relative error with respect to exact QMC benchmark data. The inconsistent incorporation of the ALDA model kernel leads to a deterioration compared to RPA for all depicted $q$.
}
\end{figure}

Let us conclude this analysis of the static density response of warm dense hydrogen by comparing our new approach to the current state-of-the-art. In Fig.~\ref{fig:H_LDA_rs4}, we again consider hydrogen at $r_s=4$ and $\theta=1$, and the black squares, blue circles, and orange triangles show the QMC, LDA, and PBE results from Fig.~\ref{fig:H_rs2_4}. In addition, the green crosses have been obtained following the standard paradigm within LR-TDDFT, that is, computing the reference function $\chi_0(\mathbf{q},\omega)$ in the limit of $\omega\to0$ on the basis of the KS-orbitals from a DFT simulation of the unperturbed system using the LDA functional. Both the KS-response function and the corresponding RPA are de-facto uncontrolled approximations. In practice, the green crosses are accurate for small $q$, but lead to a substantial deterioration in the accuracy for $q\gtrsim2q_\textnormal{F}$ compared to the LDA evaluation proposed in the present work; this can be seen particularly well in the top panel showing a the relative deviation to the exact QMC reference data.
Even worse, including the widely used ALDA model as the XC-kernel (red crosses)---a standard practice within LR-TDDFT~\cite{marques2012fundamentals,ullrich2012time}---\emph{actually increases the systematic errors} for all $q$. This constitutes an unambiguous demonstration of the practical impact of the inconsistent combination of a KS-response function with an XC-kernel from a different model, which is overcome by our new approach.

\textbf{Discussion.}
We have presented a new, formally exact framework  to compute the material specific electronic static XC-kernel within DFT, and without any additional external input apart from the usual XC-functional. 
% The new framework opens up the unprecedented possibility to compute the electronic XC-kernel of arbitrary materials; even complex atomic mixtures \cite{doi:10.1063/5.0076692, PhysRevB.85.235438, https://doi.org/10.1002/pssa.201001116} do not pose a fundamental challenge to our approach. 
Our methodology provides access to the static XC kernel across all rungs of Jacob's ladder, including promising hybrid functionals~\cite{doi:10.1063/1.1564060}. The analysis of the XC-kernel $K_\textnormal{xc}(\mathbf{q})$ on the basis of a particular functional can give valuable insights to guide new developments, as we have demonstrated for the case of WDM.
Beyond WDM, the presented framework will have a strong impact on a number of research fields within physics, chemistry, and related disciplines. Indeed, the XC-kernel is the key ingredient to a host of practical applications, such as the construction of electronically screened effective potentials~\cite{ceperley_potential,zhandos2,quantum_theory}, the incorporation of correlation effects into quantum fluid dynamics~\cite{Murillo,zhandos_pop18,Moldabekov_SciPost_2022},
and the estimation of the energy loss characteristics of high-energy density plasmas~\cite{GERICKE1996241,Moldabekov_PRE_Tlek_2020,PhysRevLett.121.145001}.
A particularly important example is given by the interpretation of XRTS experiments within the widely used Chihara approximation~\cite{Chihara_1987,kraus_xrts}. Moreover, the fluctuation--dissipation theorem gives a direct relation between the thus obtained density response and the dynamic structure factor $S_{ee}(\mathbf{q},\omega)$. % and, in this way, opens up the enticing possibility to
Consequently, our new framework for the XC-kernel opens up the enticing possibility to obtain the static structure factor $S_{ee}(\mathbf{q})$---the Fourier transform of the pair correlation function $g_{ee}(\mathbf{r})$---of two electrons \emph{exclusively within DFT and without any additional external input} apart from the usual XC-functional of standard DFT.

%%%%%%%%%%%%%%%%%%%%%%%%%%%%%%%%%%%%%%%%%%%%%%%%%%%%%%%%%%%%%%%%%%%%%%%%%%%%%%%%
% Data
%%%%%%%%%%%%%%%%%%%%%%%%%%%%%%%%%%%%%%%%%%%%%%%%%%%%%%%%%%%%%%%%%%%%%%%%%%%%%%%%
% \section*{Data Availability}
% The data supporting the findings of this study are available on the Rossendorf Data Repository (RODARE)~\cite{data}.

% \section*{Acknowledgments}
\begin{acknowledgements}We gratefully acknowledge helpful comments by K.~Burke (UC Irvine).
This work was partially supported by the Center for Advanced Systems Understanding (CASUS) which is financed by Germany’s Federal Ministry of Education and Research (BMBF) and by the Saxon state government out of the State budget approved by the Saxon State Parliament. D.B.~acknowledges support by the Polish National Science Center (NCN) under grant No. 2019/33/B/ST9/03059.

\end{acknowledgements}

%%%%%%%%%%%%%%%%%%%%%%%%%%%%%%%%%%%%%%%%%%%%%%%%%%%%%%%%%%%%%%%%%%%%%%%%%%%%%%%%
% Bibliography
%%%%%%%%%%%%%%%%%%%%%%%%%%%%%%%%%%%%%%%%%%%%%%%%%%%%%%%%%%%%%%%%%%%%%%%%%%%%%%%%
\bibliography{main}

\end{document}